\documentclass[fleqn,usenatbib]{mnras}

\usepackage{newtxtext,newtxmath}
\usepackage{bm}
\usepackage[T1]{fontenc}

\DeclareRobustCommand{\VAN}[3]{#2}
\let\VANthebibliography\thebibliography
\def\thebibliography{\DeclareRobustCommand{\VAN}[3]{##3}\VANthebibliography}

\usepackage{graphicx}	
\usepackage{amsmath}	

\title[Plasma suppression effect in FRBs not important]{The plasma suppression effect can be ignored in realistic FRB models invoking bunched coherent radio emission}

\author[Y. Qu, B. Zhang and P. Kumar]{
Yuanhong Qu$^{1,2}$\thanks{E-mail: yuanhong.qu@unlv.edu}
,
Bing Zhang$^{1,2}$\thanks{E-mail: bing.zhang@unlv.edu}
and
Pawan Kumar$^{3}$\thanks{E-mail: pk@astro.as.utexas.edu}
\\
$^{1}$Nevada Center for Astrophysics, University of Nevada, Las Vegas, NV 89154\\
$^{2}$Department of Physics and Astronomy, University of Nevada Las Vegas, Las Vegas, NV 89154, USA\\
$^{3}$Department of Astronomy, University of Texas at Austin, Austin, TX 78712, USA
}

\date{}
\pubyear{2022}

\begin{document}
\label{firstpage}
\pagerange{\pageref{firstpage}--\pageref{lastpage}}
\maketitle

\begin{abstract}
One widely discussed mechanism to produce highly coherent radio emission of fast radio bursts (FRBs) is coherent emission by bunches, either via curvature radiation or inverse Compton scattering (ICS). It has been suggested that the plasma oscillation effect can significantly suppress coherent emission power by bunches.
We examine this criticism in this paper. The suppression factor formalism was derived within the context of radio pulsars in which radio waves are in the low-amplitude, linear regime and cannot directly be applied to the large-amplitude, non-linear regime relevant for FRBs. Even if one applies this linear treatment, plasma suppression is not important for two physical reasons. First, for an efficient radiation mechanism such as ICS, the required plasma density is not high so that a high-density plasma may not exist. 
Second, both bunched coherent mechanisms demand that a large global parallel electric field ($E_\parallel$) must exist in the emission region in order to continuously inject energy to the bunches to power an FRB. In order to produce typical FRB duration via coherent curvature or ICS radiation, a parallel electric field must be present to balance the acceleration and radiation back-reaction. The plasma suppression factor should be modified with the existence of $E_\parallel$.  We show that the correction factor for curvature radiation, $f_{\rm cur}$, increases with $E_\parallel$ and becomes 1 when $E_\parallel$ reaches the radiation-reaction-limited regime. We conclude that the plasma suppression effect can be ignored for realistic FRB emission models invoking bunched coherent radio emission.
\end{abstract}




\section{Introduction}
The emission mechanism of fast radio bursts (FRBs) \citep{Lorimer07,Petroff19,Cordes19} is still puzzling since their discovery. The brightness temperature of an FRB source can be estimated as 
\begin{equation}
T_{\rm b}\simeq\frac{S_{\nu,p}D_{\rm A}^2}{2\pi k_{\rm B}(\nu\Delta t)^2}\simeq(1.2\times10^{36} \ {\rm K}) \ D_{\rm A,28}^2(S_{\nu,p}/{\rm Jy})\nu_9^{-2}\Delta t_{-3}^{-2},
\end{equation}
where $k_{\rm B}$ is Boltzmann constant, $S_{\nu,p}$ is specific flux at the peak time, $\nu$ is observing frequency, $\Delta t$ is variability timescale and $D_{\rm A}$ is angular distance of the source. Here the convention $Q_n=Q/10^n$ is adopted in cgs units. In contrast, the maximum brightness temperature of an incoherent emitting source may be estimated as \citep{Zhang22}
\begin{equation}
T_{\rm b,incoh}\simeq\Gamma_b\gamma_{\rm th} m_e c^2/k_{\rm B}\simeq(5.9\times10^{13} \ {\rm K}) \ \Gamma_{b,2}\gamma_{\rm th,2},
\end{equation}
where $\Gamma_b$ is the bulk Lorentz factor of the emitter towards observer, $m_e$ is the electron mass and $\gamma_{\rm th}$ is the thermal Lorentz factor of electrons. Both $\Gamma_b$ and $\gamma_{\rm th}$ are normalized to 100.
The extremely high brightness temperature of FRBs implies that FRB emission mechanisms must be coherent. The recent detection of the FRB-like event, FRB 200428 \citep{Bochenek20a,CHIME20a} in association with an X-ray burst \citep{CKLi2021,Mereghetti20} from the Galactic magnetar SGR 1935+2154 suggested that magnetars are the sources of at least some FRBs. 

The FRB radiation models generally include two classes \citep{Zhang2020nature}: pulsar-like models invoking emission inside or slightly outside of the magnetosphere \citep{Kumar17,Yang&Zhang2018,Kumar&Bosnjak20,Lu20,Lyubarsky2020,Wadiasingh20,YangZhang21,Zhang22,QKZ22} and GRB-like models invoking relativistic shocks far away from the magnetar \citep{waxman2017,Metzger17,Metzger19,Beloborodov17,Beloborodov20,Margalit20}. 
This paper mainly concerns the former type of model. In particular, a widely discussed radiation mechanism for pulsar-like models is coherent curvature radiation by bunches \citep{Katz14,Kumar17,Yang&Zhang2018,Lu20,Cooper21,Wang21}. Recently, \cite{Zhang22} suggested  an alternative mechanism in terms of bunched coherent inverse Compton scattering (ICS) off low frequency electromagnetic waves generated in the near-surface region of the magnetar triggered by crustal-quake-induced magnetospheric oscillations.

Charged bunches, especially those for curvature radiation, may be surrounded by a dense plasma within the magnetosphere \citep{GJ1969} and be formed through two-stream instabilities. Plasma oscillations work against radiation by bunches, which would suppress the emission power. \cite{Gil04} considered a point-like charge moving relativistically in a circularly curved magnetic field with infinite strength, surrounded by a uniform cold plasma oscillating in response of bunch emission and found a suppression factor of curvature radiation, which is $\ll 1$. Assuming that the FRB emission originates from one bunch, \cite{Lyubarsky21} calculated the value of the suppression factor in the context of FRBs and found that it is a very small value, i.e. $f_{\rm cur}^{\rm Gil} \sim 10^{-10}$, suggesting that the emitted power is about 10 orders of magnitude smaller than the vacuum case \citep{Rybicki1979,Jackson1998}. \cite{Lyubarsky21} therefore suggested that coherent curvature radiation is not a competitive mechanism to power FRB emission.

In this paper, we critically study the issue raised by \cite{Lyubarsky21} based on the \cite{Gil04} formalism. After reviewing the curvature and ICS coherent mechanisms in Section \ref{sec:2}, we point out several issues of the \cite{Gil04} formalism in \ref{sec:issues}.  Besides the obvious flaw of applying a linear treatment to the non-linear regime relevant to FRBs, there are two issues of applying the \cite{Gil04} suppression factor to FRB problems. The first is that for an efficient coherent mechanism such as ICS, a bunch can be realized by charge density fluctuations without the need of a dense plasma surrounding the bunch (Sect. \ref{sec:xi}). The second is that in order to maintain the high radiation power of the bunches there must exist an $E_\parallel$ in the emission region. The suppression factor of \cite{Gil04} does not apply in this case (Sect. \ref{sec:E||}). In Section \ref{sec:f}, we derive a more general suppression factor for coherent curvature radiation with the presence of $E_\parallel$ and prove that $f_{\rm cur}$ approaches unity at the radiation reaction limit. The conclusions are presented in Section \ref{sec:conclusions} with some discussion.

\section{Coherent emission by bunches}\label{sec:2}
In general, a charged bunch includes a large number of net charges, $Q= \pm N_{e,b} e$ emitting in phase, where $N_{e,b}$ is the total number of net charges in units of the charge unit $e$, with the $+$ and $-$ signs denoting the positron and electron bunches, respectively. Since the emission power scales as $Q^2$, in the following discussion we do not differentiate the species of the charges and adopt the $+$ sign throughout the paper. This number depends both on net charge density and coherent volume, and the latter is usually related to the wavelength of the emitted waves \citep[e.g.][]{Kumar17,Yang&Zhang2018}. The net charge density usually scales with the Goldreich Julian density in the magnetosphere \citep{GJ1969}.

It is necessary to differentiate two scaling parameters to describe the charged particle number density. A commonly defined one is the  pair multiplicity $\xi=n_{\rm tot}/n_{\rm GJ}$, which describes the production of electron-positron pairs. This parameter gives a good description of the plasma effect, but may not describe the net charge density in the bunch, since the charges of the produced electrons and positrons cancel out each other. The second parameter that is relevant to bunched coherent emisssion is the net charge factor $\zeta=n_{\rm net}/n_{\rm GJ}$. 
In order to increase $\zeta$, one needs to redistribute the pair charges through some mechanisms such as two-stream instabilities \citep[e.g.][]{Melikidze2000}. Therefore a higher $\zeta$ requires an even higher $\xi$. Models invoking a large $\zeta$ (e.g. the bunched curvature radiation as discussed below) would require a large $\xi$, so that the plasma suppression effect would be more important. In almost all previous works (but see \citealt{Zhang22}), the two parameters $\xi$ and $\zeta$ were not differentiated and were usually denoted as $\xi$.

\subsection{Coherent curvature radiation by bunches}\label{sec:curvature}

A commonly discussed FRB emission mechanism is coherent curvature radiation by bunches \citep{Kumar17,Yang&Zhang2018,Lu20}.
Relativistic bunches could be formed via two-stream instabilities in a dynamical magnetosphere, possible related to the propagation of low-frequency Alf\'ven waves triggered by the crustal deformation on the magnetar surface. The bunches move relativistically along the curved local magnetic field lines to produce coherent curvature radiation. For simplicity, the charged bunches are considered as point charges. In order to produce the 1-GHz radio waves, one requires
\begin{equation}
\Gamma_{\rm cur}\simeq241(\rho_8\nu_9)^{1/3},
\end{equation}
where $\rho$ is the curvature radius of the magnetic field line. { One important feature for the bunched coherent mechanism is that the emission power of $N_{e,b}$ leptons in a bunch is $N^2_{e,b}$ times of the emission power of individual particle. This makes the cooling time of the bunches much shorter than milliseconds. As a result, in order to continuously power a high-luminosity FRB lasting for milliseconds, a $E_\parallel$ in the emission region is needed to continuously supply energy to the bunches. }
The balance between $E_\parallel$ acceleration and curvature radiation cooling requires \citep{Kumar17}
\begin{equation}
N_{e,b}eE_{\parallel,\rm cur} c=N_{e,b}^2\frac{2e^2c\Gamma_{\rm cur}^4}{3\rho^2} \ \Rightarrow \ E_{\parallel,\rm cur}=\frac{2e\Gamma_{\rm cur}^4N_{e,b}}{3\rho^2}.
\end{equation}
The parallel electric field can be calculated as
\begin{equation}
E_{\parallel,\rm cur}\simeq(1.1\times10^{4} \ {\rm{esu}}) \ N_{e,b,20} \rho_8^{-2/3}\nu_9^{4/3},
\label{eq:Ecuv}
\end{equation}
where
\begin{equation}
N_{e,b}=n A l_{\parallel}\simeq10^{20} \ \zeta_{2}B_{\star,15}P^{-1}\hat{r}_{2}^{-3}\nu_9^{-3}
\end{equation}
is the total number of net charges in one bunch, $A=\pi(\gamma\lambda)^2=(7.6\times10^8 \ {\rm cm}^2) \ \nu_9^{-2}$ is the cross section of a bunch \citep[e.g.][]{Kumar17}, $l_{\parallel}\simeq\lambda$ is the longitudinal scale of the bunch, $\lambda = (30 \ {\rm cm}) \ \nu_9^{-1}$ is the characteristic wavelength of the FRB emission, and the GJ density is
\begin{equation}
n_{\rm GJ}\sim \frac{\Omega B}{2\pi ec}=\frac{B_{\star}\Omega}{2\pi ec}\left(\frac{r}{R_{\star}}\right)^{-3}\simeq(7\times10^{7} \ {\rm cm^{-3}}) \ B_{\star,15}P^{-1}\hat{r}_{2}^{-3}
\end{equation}
at a normalized radius $\hat r = r/R_{\star} = 100 \ \hat r_2$ ($R_{\star}$ is the neutron star radius). If one neglects the plasma suppression effect, the total emission power of $N_b$ bunches can be estimated as
\begin{equation}
\begin{aligned}
P_{\rm vac,cur}&=N_bN_{e,b}^2\Gamma_{\rm cur}^2\frac{2e^2\Gamma_{\rm cur}^4c}{3\rho^2}\\
&\simeq(9.0\times10^{37} \ {\rm erg} \ {\rm s^{-1}}) \
N_{b,8}N_{e,b,20}^2\Gamma_{\rm cur,2.38}^6\rho_8^{-2}.
\end{aligned}
\end{equation}
This emission is beamed within a $1/\Gamma_{\rm cur}$ cone. For an observer looking down into the relativistic beam, the observed isotropic luminosity is a factor $\sim \Gamma_{\rm cur}^2$ larger, i.e.
\begin{equation}
\begin{aligned}
L_{\rm vac,cuv}&=P_{\rm vac,cuv} \Gamma_{\rm cur}^2 \\
&= (5.2 \times 10^{42} \ {\rm erg \ s^{-1}}) \ N_{b,8}N_{e,b,20}^2\Gamma_{\rm cur,2.38}^8\rho_8^{-2}.
\end{aligned}
\label{eq:L}
\end{equation}
Note that even if the plasma suppression effect is not considered, $\zeta \sim 10^2$ and $N_b \sim 10^{8}$ are needed to reach the desired isotropic luminosity of a bright FRB. 

In the literature, the charges producing coherent curvature radiation are sometimes delineated as one giant charge surrounded by a plasma \citep[e.g.][]{Gil04,Lyubarsky21}. Since $\zeta \gg 1$ is required, one has $\xi \gg 1$ and the plasma suppression effect should be considered. However, whatever the bunching mechanism is, it is hard to justify that only one bunch is generated as assumed by \cite{Lyubarsky21}. Very generally, the observed total luminosity should be the incoherent superposition of a total of $N_b$ bunches (see Eq.(\ref{eq:L})).

\subsection{Coherent inverse Compton scattering by bunches}\label{sec:ICS}
The general picture of the ICS mechanism is the following \citep{Zhang22}: A low frequency electromagnetic wave with angular frequency $\omega_{i}$ would be excited from the surface of the magnetar due to the {crust cracking and plasma oscillations} near the surface of the magnetar. The waves could propagate freely within the magnetosphere if allowed by the dispersion relation { and} would be upscattered by relativistic bunches to produce FRBs.
\cite{Zhang22} showed (in Appendix) that in the case of $\vec k \parallel \vec B$ ($\vec k$ is the low-frequency wave number vector and $\vec B$ is the local magnetic field vector), both eigenmodes (L-mode and R-mode) of the waves can propagate freely. In the case of $\vec k \perp \vec B$, only the X-mode ($\vec E_w \perp \vec B$, where $E_w$ is the electric field vector of the low-frequency wave) can propagate, while the O-mode ($\vec E_w \parallel \vec B$) propagation is forbidden. He did not derive dispersion relation in most general oblique case ($0 < \left< \vec k, \vec B\right> < \pi/2$, where $\left< \vec k, \vec B\right>$ is the angle between $\vec k$ and $\vec B$). Since in the parallel case both eigen-modes are ``extraordinary'' ($\vec E_w \perp \vec B$) and are free to propagate, he more generally stated that the ``X-mode'' of the waves can propagate freely. 

In the more general oblique case, the two eigen modes are defined as the mode (1), where $\vec k$ is parallel to the $(\vec k, \vec B)$ plane, and mode (2), where $\vec k$ is perpendicular to the $(\vec k, \vec B)$ plane. In the literature, mode (1) is also called the ``O-mode'' while mode (2) is also called the ``X-mode''. However, {this} terminology is more relevant in the quasi-perpendicular case, i.e. when $\vec k$ and $\vec B$ are nearly perpendicular to each other. In the quasi-parallel case ($\vec k$ and $\vec B$ are nearly parallel), both modes (1) and (2) are nearly extraordinary and the so-called O-mode can also propagate. The mode (2) (X-mode) can propagate freely in any case. { In Appendix \ref{sec:dispersion}, we have treated the dispersion relation of waves with an arbitrary $\theta=\left<\vec k, \vec B \right>$ angle. For the (2) mode (X-mode), one can prove that the wave can propagate freely in the entire $\omega-\theta$ plane. For the (1) mode (O-mode), the situation is more complicated. We have calculated the cutoffs and resonances and marked the forbidden region for wave propagation in Figure \ref{fig:fre}. As shown by the figure, mode (1) low-frequency waves can propagate in essentially all $\left< \vec k, \vec B\right>$ angles except $\left< \vec k, \vec B\right> \simeq \pi/2$. For the outgoing scattered FRB waves with GHz frequency, mode (1) can propagate if $\left< \vec k, \vec B\right> \lesssim 0.33$.  In the ICS model of FRBs, the low-frequency waves have an incident angle $< \pi/2$ and the outgoing FRB waves travel with a direction nearly parallel to the local $\vec B$ field. As a result, mode (1) can travel for both the incoming and outgoing radio waves.}

In the rest frame of electron, the Compton scattering cross section for the two modes of photons in a strong magnetic field is given by \citep{Herold1979,Xia1985}
\begin{equation}\label{eq:Omode}
\sigma'(1)=\sigma_{\rm T}\left\{\sin^2\theta_i'+\frac{1}{2}\cos^2\theta_i'\left[\frac{\omega'^2}{(\omega'+\omega'_B)^2}+\frac{\omega'^2}{(\omega'-\omega'_B)^2}\right]\right\},
\end{equation}
and
\begin{equation}\label{eq:Xmode}
\sigma'(2)=\frac{\sigma_{\rm T}}{2}\left[\frac{\omega'^2}{(\omega'+\omega'_B)^2}+\frac{\omega'^2}{(\omega'-\omega'_B)^2}\right],
\end{equation}
where $\sigma'(1)$ and $\sigma'(2)$ denote the cross sections that the photons are scattered to modes (1) and (2), respectively, and $\theta'_i$ is the incident angle $\left<\vec k', \vec B'\right>$ of the incoming waves. The two terms within the braces in Eq.(\ref{eq:Omode}) denote the contributions of (1)-to-(1) and (2)-to-(1) scatterings, respectively, while the two terms within the square brackets in Eq.(\ref{eq:Xmode}) denote the contributions of (1)-to-(2) and (2)-to-(2) scatterings, respectively. For typical parameters in a magnetar magnetosphere, the (1)-to-(1) term is much greater than the other three terms (which is of the order of $(\omega'/\omega'_B)^2 \ll 1$). Since the mode (1) of both incoming and outgoing radio waves can propagate, the cross section is dominated by the $\sigma'=\sigma_{\rm T} \sin^2\theta'_i$ term, as suggested by \cite{Zhang22}.

In the lab frame, the ICS cross section is \citep{Qiao98,Zhang22}
\begin{equation}
    \sigma \simeq \sigma'(1) (1-\beta \cos\theta_i) = \sigma_{\rm T} \frac{\sin^2 \theta_i}{\Gamma_{\rm ICS}^2 (1-\beta\cos\theta_i)} = f(\theta_i) \sigma_{\rm T}/\Gamma_{\rm ICS}^2,
\end{equation}
where $\theta_i$ is the incident angle  $\left<\vec k, \vec B\right>$ of the incoming waves in the lab frame, $\Gamma_{\rm ICS}$ is the Lorentz factor of the changed bunch, and  $f(\theta_{i})=\sin^2\theta_{i}/(1-\beta\cos\theta_{i})$ is a factor of order unity.

In order to produce 1-GHz radio waves through the ICS mechanism one requires
\begin{equation}\label{GammaICS}
\Gamma_{\rm ICS}\simeq316 \ \nu_{\rm 9}^{1/2}\nu_{ i,4}^{-1/2}(1-\beta\cos\theta_{i})^{-1/2},
\end{equation}
where $\theta_{i}$ is the angle between the incident photon momentum and the electron momentum, $\nu_{i}$ and $\nu$ denote the frequencies of the incident and scattered waves, respectively. 
The balance between acceleration and ICS radiation cooling requires
\begin{equation}
E_{\parallel,\rm ICS}\simeq(8.6\times10^{6} \ {\rm esu}) \ \zeta_2\nu_9^{-3}\Gamma_{\rm ICS,2.5}^{2}B_{\star,15}P^{-1}f(\theta_{i})\delta B_{0,6}^2\hat{r}_3^{-5},
\label{eq:EICS}
\end{equation}
where $\hat r = r/R$ is the emission radius normalized to the radius of the neutron star $R$.

The ICS power of individual electrons is (7-8) orders of magnitude higher than that of curvature radiation. As a result, the ICS radiation power (without plasma suppression effect) of $N_b$ relativistic bunches can be estimated as \citep{Zhang22}
\begin{equation}
\begin{aligned}
P_{\rm vac, ICS}&\simeq N_bN_{e,b}^2\Gamma_{\rm ICS}^2\frac{4}{3}\Gamma_{\rm ICS}^2\sigma cU_{\rm ph}\\
&\simeq(2.1\times10^{38} \ {\rm erg \ s^{-1}}) \ N_{b,5}N_{e,b,18}^2\Gamma_{\rm ICS,2.5}^2f(\theta_i)\delta B_{0,6}^2\hat{r}_2^{-2},
\end{aligned}
\end{equation}
where {the emission power of $N_{e,b}$ particles is also enhanced by a factor of $N_{e,b}^2$ in the coherent ICS model}, $\sigma \sim f(\theta_i) \sigma_{\rm T} / \Gamma_{\rm ICS}^2 \sim 10^{-5} \sigma_{\rm T} \Gamma_{\rm ICS,2.5}^2$ is the scattering cross section,
$\delta B_{0}$ is the initial amplitude of the low-frequency electromagnetic waves at the surface, 
and
\begin{equation}
N_{e,b}=n A l_{\parallel}\simeq10^{18} \ \zeta B_{\star,15}P^{-1}\Gamma_{2.5}^2\hat{r}_{2}^{-3}\nu_9^{-3}
\end{equation}
is adopted, where $l_\parallel \sim \lambda$ and $A \sim \pi (\Gamma\lambda)^2$ with $\lambda$ being the wavelength of the FRB waves. 
The observed isotropic luminosity is
\begin{equation}
\begin{aligned}
L_{\rm vac,ICS}&=P_{\rm vac,ICS}\Gamma_{\rm ICS}^2\\
&\simeq(2.1\times10^{43} \ {\rm erg \ s^{-1}}) \ N_{b,5}N_{e,b,18}^2\Gamma_{\rm ICS,2.5}^4f(\theta_i)\delta B_{0,6}^2\hat{r}_2^{-2}.
\end{aligned}
\end{equation}
Note that $\zeta \sim 1$ can be adopted for such a model, with a much smaller total number of bunches, $N_b \sim 10^5$, to produce the characteristic FRB luminosity. The bunches in the coherent ICS model are merely density fluctuations. One can in principle produce FRB emission with a fluctuating Goldreich-Julian outflow without the association of a pair plasma, i.e. $\xi \sim \zeta \sim 1$. As a result, the plasma suppression effect could be neglected for this model\footnote{Other plasma effects, e.g. the plasma nature of the low-frequency waves, may be considered. However, the general feature of the model, i.e. the high emission power of the process, should be retained as argued by \cite{Zhang22}.}.

We note that induced Compton scattering is not considered within magnetospheric FRB models.
This is because in order to excite induced Compton scattering, electrons should be allowed to jump between two energy levels with an interval defined by the typical frequency of the FRB emission, $\nu_{\rm FRB} = (10^9 \ {\rm Hz}) \ \nu_9$, On the other hand, electrons can only stay at Landau levels in a strong magnetic field. The minimum energy between two Landau levels is $E_{\rm min} =\hbar{eB}/(m_ec)$. The condition for induced Compton scattering is $E_{\min} < h \nu_{\rm FRB}$, which gives $B < B_c = (357 \ {\rm G}) \ \nu_9$. In the FRB emission region (typically $\hat r \sim 100)$, one has $B \sim (10^9 \ {\rm G}) \ B_{\star,15} \hat r_2^{-3} \gg B_c$ for a magnetar with a surface magnetic field $B_{\star} = (10^{15} \ {\rm G}) \ B_{\star,15}$. Therefore, induced Compton scattering is not allowed within a magnetar magnetosphere.

\section{Plasma suppression effect in FRBs: a critique}\label{sec:issues}

\subsection{General discussion}\label{sec:general}

Within the context of radio pulsars, considering curvature radiation emitted by a charged bunch with bulk Lorentz factor $\Gamma$ surrounded by a cold plasma moving with a Lorentz factor $\gamma < \Gamma$, \cite{Gil04} introduced a suppression factor
\begin{equation}\label{eq:f_Gil}
f_{\rm cur}^{\rm Gil}=\frac{1}{4\gamma}\left(\frac{\omega_c'}{\omega_p'}\right)^2\left(1-\frac{\gamma^2}{\Gamma^2}\right)^2,
\end{equation}
where $\omega_c'=c\Gamma^3/(\rho\gamma)$ is the characteristic frequency of curvature radiation in the comoving frame and
$\rho$ is the curvature radius.
One can see that the suppression degree is proportional to $\sim(\omega_c'/\omega_p')^2/\gamma$,
which could be extremely small if the plasma is dense enough. Also, at $\gamma = \Gamma$, $f_{\rm cuv}^{\rm Gil}=0$, suggesting that the emission is completely suppressed if the plasma and the bunch move with the same Lorentz factor. \cite{Lyubarsky21} applied this formula to FRBs by assuming that the FRB emission originates from one macro bunch, and argued that $f_{\rm cuv}^{\rm Gil}$ is of the order of $10^{-10}$, so that coherent curvature radiation by bunches cannot account for the FRB emission. 

However, there are several issues of applying Eq.(\ref{eq:f_Gil}) to the FRB problem.

First of all, Eq.(\ref{eq:f_Gil}) was derived within the pulsar context by assuming that the magnetic field $B$ has an infinite strength. The plasma oscillations are introduced as a small perturbation so that the treatment was fully in the linear regime. In the case of an FRB, the waves are in the highly non-linear regime, characterized by a nonlinear parameter defined as
\begin{equation}
a=\frac{e E_w}{m_ec\omega}=\frac{e L_{\rm frb}^{1/2}}{m_e c^{3/2}\omega  r}\simeq1.6\times10^4 \ L_{\rm frb,42}^{1/2}\nu_9^{-1} \hat r_3^{-1} \gg 1.
\end{equation}
In the FRB emission region, the wave electric field strength is even stronger than the magnetic field strength and electrons are accelerated to high Lorentz factors \citep{QKZ22}. The treatment of \cite{Gil04} no longer applies. One cannot simply use Eq.(\ref{eq:f_Gil}) to quantify the plasma suppression effect.

The full treatment of the plasma suppression effect in the non-linear regime is beyond the scope of the current paper. However, even within the framework of linear treatment of \cite{Gil04}, there are still two issues against the argument of \cite{Lyubarsky21}. 

First, \cite{Lyubarsky21} argued for a very small $f_{\rm cur}^{\rm Gil}$ by assuming  $\omega'_c \ll \omega'_p$, i.e. there exists a very dense plasma in the emission region. However, this requirement is not necessarily needed, especially if the radiation mechanism is ICS. Second, in the FRB emission region there must exist a large $E_\parallel$, which will separate electron positron pairs and affect the plasma suppression factor. We discuss these two effects in the following.

\subsection{Plasma density and the required multiplicity factor $\xi$}\label{sec:xi}
The first physical condition for significant plasma suppression is the existence of a dense pair plasma, i.e. the pair-multiplicity parameter $\xi \gg 1$. One can make this condition more quantitative. The physical reason that the plasma will suppress emission is that oscillations of the emitted waves with angular frequency $\omega$ would induce plasma oscillations with the angular frequency 
\begin{equation}
\begin{aligned}
\omega_p=\sqrt{\frac{4\pi e^2n}{m_e}}&=\sqrt{\frac{2e\xi B_{\star}\Omega}{m_ec}}\left(\frac{r}{R_{\star}}\right)^{-3/2}\\
&\simeq(4.7\times10^9 \ {\rm rad \ s^{-1}}) \ \xi_2^{1/2}B_{\star,15}^{1/2}P^{-1/2}\hat{r}_{2}^{-3/2}\\
&\simeq(1.5\times10^7 \ {\rm rad \ s^{-1}}) \ \xi^{1/2}B_{\star,15}^{1/2}P^{-1/2}\hat{r}_{3}^{-3/2},
\end{aligned}
\end{equation}
where the last two expressions have made use of the typical parameters for the curvature radiation and ICS mechanisms, respectively.

If $\omega_p'/\omega'=\omega_p / \omega > 1$, plasma oscillations would be significant, which would produce a strong screening electric field to suppress the radiation of the bunch. On the other hand, if $\omega_p'/\omega'=\omega_p / \omega \ll 1$ is satisfied, the response of the plasma is very slow compared with the radiated waves. The suppression effect is therefore negligible. As a result, the first physical condition for plasma suppression can be expressed as
\begin{equation}
    \omega_p' > \omega',
    \label{eq:omegacond}
\end{equation}
which can be translated to
\begin{equation}
\xi>\frac{m_ec\omega^2}{2eB_{\star}\Omega}\left(\frac{r}{R_{\star}}\right)^3\simeq 1.8\times10^2 \ \nu_9^2B_{\star,15}^{-1}P\hat{r}_2^3.
\end{equation}
As discussed in Section \ref{sec:2},  $\xi > \zeta \sim 100$ is required by the curvature radiation model. So plasma suppression may be considered for this model. However, for the ICS model $\zeta \sim 1$ is already enough to power the observed FRB luminosity. The plasma suppression effect would not be important for the ICS model.

\subsection{Parallel electric field $E_\parallel$}\label{sec:E||}
The second physical condition for significant plasma suppression is that the bunch is surrounded by a plasma. As discussed in Sections \ref{sec:curvature} and \ref{sec:ICS}, in both coherent bunching models to produce FRB emission, a significant $E_\parallel$ is needed to continuously supply energy to the bunches in order to provide the necessary power for FRBs. Such an $E_\parallel$ tends to separate the pair plasma \citep{Yang20}. 

One may define a critical $E_{c,\parallel}$, which is the one required to separate an $e^\pm$ plasma surrounding a bunch. We consider a pair number density in a magnetar magnetosphere as
\begin{equation}
n=\frac{\xi B_{\star}\Omega}{2\pi qc}\left(\frac{r}{R_{\star}}\right)^{-3}\simeq(7\times10^{9} \ {\rm cm^{-3}}) \ \xi_{2}B_{\star,15}R_{\star,6}^3P^{-1}\hat{r}_{2}^{-3}.
\end{equation}
The mean separation distance between an electron and a positron may be estimated as $l=n^{-1/3}\simeq(5\times10^{-4} \ {\rm cm^{-3}}) \ \xi_{2}^{-1/3}B_{\star,15}^{-1/3}R_{\star,6}^{-1}P^{1/3}\hat{r}_{2}$. The critical parallel electric field needed to separate this pair may be estimated as 
\begin{equation}
E_{c,\parallel}\sim \frac{e}{l^2}\simeq(1.7\times10^{-3} \ {\rm esu}) \ \xi_2^{2/3}B_{\star,15}^{2/3}R_{\star,6}^{2}P^{-2/3}\hat{r}_2^{-2},
\end{equation}
which means that the pairs are easily separated. However, the situation is more complicated with the presence of macro bunched charges with charge $Q$. The plasma tends to be associated with the bunches through Coulomb interaction. Take the typical separation between the bunch and a charged particle as $\Delta \sim \lambda \simeq 30$ cm \citep{Melikidze2000}. The critical $E_\parallel$ needed to overcome this Coulomb force is 
\begin{equation}
E_{c,\parallel,{\rm cur}}=\frac{Q_{\rm cur}e}{\Delta^2}\simeq(5.3\times10^{7} \ {\rm esu}) \ \zeta_{2}B_{\star,15}P^{-1}\hat{r}_{2}^{-3}\nu_9^{-3},
\end{equation}
and
\begin{equation}
E_{c,\parallel,{\rm ICS}}=\frac{Q_{\rm ICS}e}{\Delta^2}\simeq(5.3\times10^{2} \ {\rm esu}) \ \zeta B_{\star,15}P^{-1}\hat{r}_{3}^{-3}\nu_9^{-3},
\end{equation}
for the curvature and ICS mechanisms, respectively, where $Q_{\rm cur/ICS}=eN_{e,b}$ is net charge of one bunch. Comparing these critical values with those required by the curvature radiation (Eq.(\ref{eq:Ecuv})) and ICS (Eq.(\ref{eq:EICS})) mechanisms, one can see that the pair plasma is bound to the bunches in the case of curvature radiation but is charge-separated in the case of the ICS mechanism. This again suggests that the plasma suppression effect is irrelevant to the ICS mechanism even if bunches are initially surrounded by a plasma. As shown in Section \ref{sec:f} below, even if $E_\parallel$ cannot fully separate charges (e.g. for the curvature radiation case), it will modify the plasma suppression factor.

\section{Plasma suppression factor with the presence of a parallel electric field}\label{sec:f}

Let us consider that a dense plasma is surrounding a charged bunch, which is radiating curvature radiation with the presence of an $E_\parallel$. The expression of $f_{\rm cur}$ would be modified. 

We start by summarizing the theoretical framework of \cite{Gil04}. Since the background magnetic field is strong in the linear treatment, charged particles can only move along magnetic field lines. One can adopt a cylindrical geometry, assign the $\theta$-direction as the direction of the magnetic field, and assume that the magnetic field has a semi-circular configuration. Since the field line is circular, the radius $\rho$ can be adopted to denote the curvature radius for curvature radiation. Next, one can divide the current density into two parts: $j_{\rm bunch}$ produced by the charged bunch moving along the magnetic field and $j_{\rm plasma}$  produced by the ambient plasma induced by the oscillating electric field of curvature radiation. The bunch current density is $j_{\rm bunch}=(QV/R)\delta(r-R)\delta({\theta-Vt/R})\delta(z)$, where $\theta$ is the azimuthal angle, $Q$ and $V$ are the total charge quantity and velocity of the bunch. The plasma current and bunch current can be excited only in the $\theta$-direction. By introducing perturbations in plasma number density and velocity, one can treat the problem in the Fourier domain, leading to Eq.(\ref{eq:f_Gil}). See the details in \cite{Gil04}\footnote{There is a typo in Eq. (A2) in the Appendix of \cite{Gil04}. It should read $\delta\tilde v=ie\tilde E_\theta/[\gamma^3m_e(\omega-sv/r)]$.}. In the following we derive a more general expression of $f_{\rm cur}$ by introducing $E_\parallel$ in the problem.

\subsection{Constant parallel electric field}
\begin{figure}
	\includegraphics[width=\columnwidth]{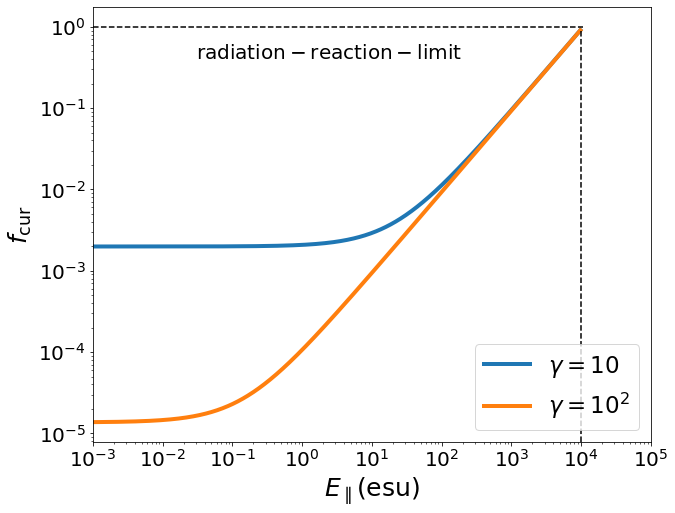}
    \caption{The suppression factor as a function of parallel electric field intensity. Following parameters are adopted: $\xi=10^2$, $B_\star=10^{15} \ \rm G$, $R_\star=10^6$ cm, $P=1$ s. The Lorentz factor of curvature bunch is $\Gamma=241$.}
\label{fig:factor}
\end{figure}

We first consider the case that $E_\parallel$ is essentially a constant. This assumption is relevant for FRBs, which requires that emission happens in the radiation-reaction-limited regime. 

The curvature radiation power of the bunch can be most generally calculated as
\begin{equation}\label{factor}
P=j_{\rm bunch}(E_\theta+E_\parallel).
\end{equation}
Here
\begin{equation}
E_\theta=\frac{Q c \Gamma^4}{6\rho^2 V \gamma}\left(\frac{c\Gamma^3}{\omega_p'\rho\gamma}\right)^2\left(1-\frac{\gamma^2}{\Gamma^2}\right)^2,
\end{equation}
is the global electric field of the system involving a bunch with charge $Q$ and velocity $V$ and the surrounding plasma without the existence of $E_\parallel$, as derived in \cite{Gil04}.
Note that similar to \cite{Gil04}, we have assumed that the plasma Lorentz factor $\gamma$ is smaller than the bunch Lorentz factor $\Gamma$ so that one can have a direct comparison between the two treatments\footnote{Strictly speaking, the assumption of $\gamma < \Gamma$ may not be always satisfied for both \cite{Gil04} and our treatment. A self-consistent, time-dependent calculation of $\gamma$ and $\Gamma$ is needed. However, the purpose of this paper is to check how $f_{\rm cur}$ is modified from $f_{\rm cur}^{\rm Gil}$ given the same input parameters.}.
Therefore, the emission power of the point-like charge can be calculated as $P=j_{\rm bunch}(E_\theta+E_\parallel)$ and the degree of suppression effect can be compared with vacuum case in Eq. (\ref{eq:f_Gil}).
For a constant parallel electric field, we provide a derivation in Appendix \ref{A} to prove that the presence of $E_\parallel$ would not affect the suppression factor term corresponding to $E_\theta$. Therefore, 
the modified total emission power should be calculated as
\begin{equation}
P=j_{\rm bunch}(E_\theta+E_\parallel)=QV\left[\frac{Qc\Gamma^4}{6\rho^2V\gamma}\left(\frac{c\Gamma^3}{\omega_p'\rho\gamma}\right)^2\left(1-\frac{\gamma^2}{\Gamma^2}\right)^2+E_\parallel\right].
\end{equation}
This gives the modified suppression factor in the presence of $E_\parallel$
\begin{equation}
\begin{aligned}
f_{\rm cur}=\frac{P}{P_{\rm cur}}&=\frac{1}{4\gamma}\left(\frac{c\Gamma^3}{\omega_p'\rho\gamma}\right)^2\left(1-\frac{\gamma^2}{\Gamma^2}\right)^2+\frac{QVE_\parallel}{P_{\rm cur}}\\
&\simeq\frac{1}{4\gamma}\left(\frac{c\Gamma^3}{\omega_p'\rho\gamma}\right)^2\left(1-\frac{\gamma^2}{\Gamma^2}\right)^2+\frac{QcE_\parallel}{P_{\rm cur}},
\label{eq:f_cur}
\end{aligned}
\end{equation}
where we have assumed $V\sim c$. One can see that Eq.(\ref{eq:f_cur}) has two terms: the first term is the same as that obtained by \cite{Gil04}, while the second term is a term related to the strength of $E_\parallel$. The value of $f_{\rm cur}$ as a function of $E_\parallel$ for a fixed $\Gamma$ value and different $\gamma$ values are presented in Fig. \ref{fig:factor}. One can see that $f_{\rm cur}$ increases with $E_\parallel$, suggesting that the presence of an $E_\parallel$ can effectively cancel the plasma suppression effect. Interestingly, at the radiation reaction limit, i.e. $QcE_\parallel=P_{\rm cur}$, the second term of $f_{\rm cur}$ becomes unity, which completely removes the plasma suppression effect. This suggests that even for a linear treatment, under the realistic condition for FRB production there is no plasma suppression effect.

\subsection{Slow-changing parallel electric field}

In reality, $E_\parallel$ is likely not a constant all the time. More realistic models \citep{Kumar&Bosnjak20} suggests that $E_\parallel$ itself is oscillating, but with a much lower frequency. 
We provide a derivation in Appendix \ref{A} to prove that such a slowly oscillating electric field would not modify the suppression factor significantly. Here we consider a specific mechanism that invokes a charge-starved Alfv\'en wave as the source of $E_\parallel$ \citep{Kumar&Bosnjak20}.
The electric field parallel to background magnetic field can be written as
\begin{equation}
E_\parallel=E_d{\rm exp}(ik_{\rm aw}z-i\omega_{\rm aw}t),
\end{equation}
where $E_d$ is the wave amplitude, $k_{\rm aw} \ll k$ and $\omega_{\rm aw} \ll \omega$ are the wave vector and circular frequency of Alfevn wave, respectively. The modified suppression factor can be written as
\begin{equation}
\begin{aligned}
f_{\rm cur}&\simeq\frac{1}{4\gamma}\left(\frac{c\Gamma^3}{\omega_p'\rho\gamma}\right)^2\left(1-\frac{\gamma^2}{\Gamma^2}\right)^2+\frac{QcE_d}{P_{\rm cur}}{\rm exp}(ik_{\rm aw}z-i\omega_{\rm aw}t)\\
&\sim \frac{1}{4\gamma}\left(\frac{c\Gamma^3}{\omega_p'\rho\gamma}\right)^2\left(1-\frac{\gamma^2}{\Gamma^2}\right)^2+\frac{QcE_d}{P_{\rm cur}},
\end{aligned}
\end{equation}
where the oscillation factor is of order of unity and can be approximately dropped out. The results are similar to the constant $E_\parallel$ case.

\section{Conclusions and Discussion}\label{sec:conclusions}
In this paper, we critically examined the suggested plasma suppression effect for bunched coherent radio emission \citep{Gil04,Lyubarsky21} within the framework of FRBs. We reached the conclusion that such a suppression effect is not important based on the following arguments:
\begin{itemize}
\item The suppression factor Eq.(\ref{eq:f_Gil}) derived by \cite{Gil04} was based on the assumption that the magnetic field has infinite strength. This is valid for radio pulsar emission but in the case of FRB emission, the radiation is so intense that the electric field strength in the waves is comparable to or even exceeds the strength of the background magnetic field. The linear perturbation treatment of \cite{Gil04} is no longer valid. A treatment of the problem in the non-linear regime is beyond the scope of this paper, but one should be cautious of using the suppression factor Eq.(\ref{eq:f_Gil}) to the FRB problems.
\item Even within the framework of the linear treatment, the plasma suppression effect can be ignored if there is no dense plasma surrounding the emitting bunch. This condition can be reached under two conditions: (1) the emission power of the bunch is high so that no dense plasma is needed to power FRB emission; (2) there is a strong external $E_\parallel$ component in the emission region so that the electron-positron pairs are separated. Both conditions can be satisfied for the coherent ICS mechanism for bunches. 
\item For coherent curvature radiation by bunches, a dense plasma may still accompany the emitting bunches. However, the existence of an $E_\parallel$ in the emission region, as is required by the bunched coherent radiation models for FRBs, would modify the suppression factor. We show that $f_{\rm cur}$ increases with $E_\parallel$ and reaches unity when $E_\parallel$ reaches the radiation reaction limited regime, as is required by the FRB model. This suggests that the bunch emission power is similar to the vacuum value and the suppression factor can be ignored. 
\end{itemize}

\section*{Acknowledgements}
We thank Rei Nishiura and Kunihito Ioka for helpful comments.
This work is partially supported by the Top Tier Doctoral Graduate Research Assistantship (TTDGRA) and Nevada Center for Astrophysics at the University of Nevada, Las Vegas. PK and YQ acknowledge NSF grant AST-2009619 that supported this work.

\section*{Data Availability}
The detailed derivations presented in this paper are available upon request.




\appendix

\section{Dispersion relation of a cold plasma in the oblique case}\label{sec:dispersion}

\begin{figure*}
\begin{center}
\begin{tabular}{ll}
\resizebox{80mm}{!}{\includegraphics[]{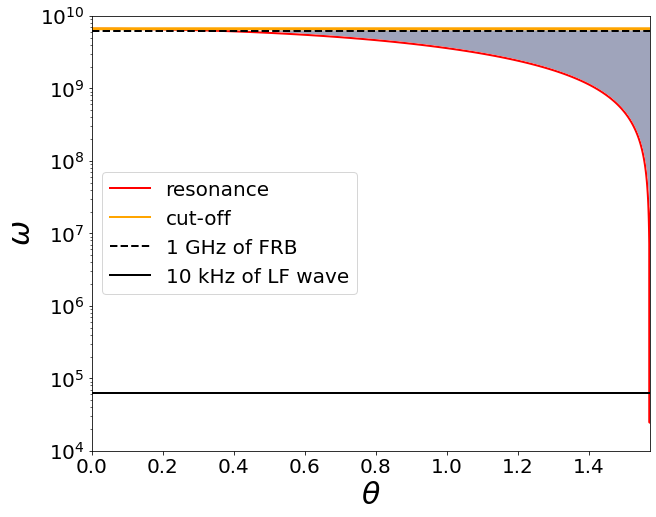}}&
\resizebox{80mm}{!}{\includegraphics[]{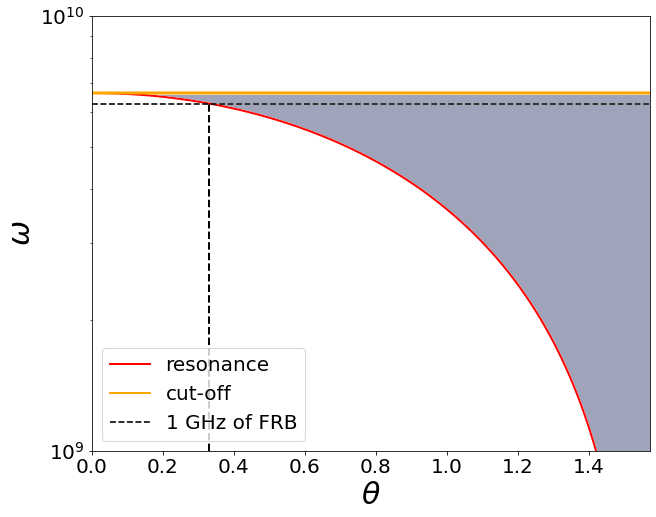}}
\end{tabular}
\caption{The resonance and cut-off frequencies for O-mode waves as a function of $\theta = \left< \vec k, \vec B \right>$ in units of radian. Following parameters are adopted: $\xi=10^2$, $B_\star=10^{15} \ \rm G$, $R_{\star}=10^6 \ \rm cm$, $P=1 \ \rm s$ and $r=10^8 \ \rm cm$. The grey zone is the forbidden region for wave propagation. The right panel is the zoom-in version of the  left panel. The O-mode low frequency waves presented in the left panel can propagate in all angles except $\theta = \pi/2$.  The scattered wave (dashed line in right panel) can propagate as long as $\theta < 0.33$ (see the perpendicular dashed line in the right panel).}
\label{fig:fre}
\end{center}
\end{figure*}

In this Appendix, we provide a detailed treatment of the dispersion relation of waves propagating in a cold plasma for the general ``oblique'' case, i.e. $\vec k$ and $\vec B$ have an arbitrary angle. For a pair plasma, the plasma frequency $\omega_p$ is defined using the number density of electrons only. If $\omega_p$ is defined using the pair density instead, all the $2\omega_p^2$ terms should be changed to $\omega_p^2$.

The general expression of the dispersion relation can be written as \citep{stix1992}
\begin{equation}
n^2=\frac{B\pm F}{2A},
\label{eq:n2}
\end{equation}
where the ``$+$'' sign denotes the X-mode (or (2) mode in the text), which means that the wave electric vector $\Vec{E}_w$ is perpendicular to $\Vec{k}-\Vec{B}$ plane, and the ``$-$'' sign denotes the O-mode (or (1) mode in the text), which means that $\Vec{E}_w$ is parallel to the $\Vec{k}-\Vec{B}$ plane. 
The three parameters $A$, $B$ and $F$ in the cold pair plasma can be written as
\begin{equation}
A=\left(1-\frac{2\omega_p^2}{\omega^2-\omega_B^2}\right)\sin^2\theta+\left(1-\frac{2\omega_p^2}{\omega^2}\right)\cos^2\theta,
\end{equation}
\begin{equation}
\begin{aligned}
B=\left(1-\frac{2\omega_p^2}{\omega^2-\omega_B^2}\right)^2\sin^2\theta+&\left(1-\frac{2\omega_p^2}{\omega^2}\right)\left(1-\frac{2\omega_p^2}{\omega^2-\omega_B^2}\right)\\
&\times\left(1+\cos^2\theta\right),
\end{aligned}
\end{equation}
and
\begin{equation}
F=\sqrt{B^2-4AC},
\end{equation}
where
\begin{equation}
C=\left(1-\frac{2\omega_p^2}{\omega^2}\right)\left(1-\frac{2\omega_p^2}{\omega^2-\omega_B^2}\right)^2,
\end{equation}
and $\theta$ is the angle between $\Vec{k}$ and $\Vec{B}$.

A radio wave can propagate if Eq.(\ref{eq:n2}) has a real solution, i.e. $0\leq n^2 < +\infty$. In order to find the allowed region in the $\omega-\theta$ plane, one needs to first find the boundaries of the allowed regions, which are defined by ``cutoffs'' (corresponding to $n=ck/\omega=0$) and ``resonances'' (corresponding to $n=ck/\omega=\infty$).

First, it is observed that $n^2=(B+F)/2A>0$ is always satisfied, so that the X-mode (or (2) mode) can propagate in all $\theta$ for all $\omega$ in the magnetosphere. Since $|F| < |B|$, $(B+F)$ always carries the same sign of $B$. Inspecting $B$ and $A$ and noticing $\omega \ll \omega_p \ll \omega_{B}$ for the problem being solved, one can prove that $B/A$ is always positive so that $(B+F)/2A$ is always positive.

The case of O-mode ((1) mode) is more complicated. The numerical results of the cutoff ($B-F=0$) and  resonance ($A=0$) frequency are presented in the left panel of Fig. \ref{fig:fre}. The forbidden region is marked in grey. The two critical frequencies of interest, i.e. $\omega_i = (2\pi) 10^4 \ {\rm rad \ s^{-1}}$ for the low-frequency waves in the ICS model and $\omega = (2\pi) 10^9 \ {\rm rad \ s^{-1}}$ for the FRB waves, are plotted for comparison. One can see that the O-mode low-frequency waves can propagate freely essentially in all angles except strictly $\theta=\pi/2$ (the perpendicular case).
For the FRB waves, since $\omega$ and the  cutoff frequency are very close, we zoom in in the right panel of Fig. \ref{fig:fre}. One can see that the O-mode FRB waves can propagate freely at $\theta \lesssim 0.3$. This is the case for the ICS model, since the outgoing waves are expected to align with the relativistic electron moving direction, which is the $\Vec B$ vector direction in the emission region. During the propagation process $\theta$ would not go to very large values, especially with the possible alignment of $\vec k$ and $\vec B$ due to the strong radiation pressure of the FRB waves \citep{QKZ22}.

\section{Suppression factor in the presence of parallel electric field}\label{A}
In this Appendix, we derive the modified suppression factor in the presence of a parallel electric field $E_\parallel$. 

We consider the plasma motion under the influence of two electric fields: $E_\theta$ originated from the plasma oscillations and curvature radiation and a global external $E_{\parallel}$ due to the large scale electrodynamics. The relativistic motion equation of a single charged particle is
\begin{equation}\label{motion}
m_e\gamma^3\frac{dv}{dt}=e(E_\theta+E_\parallel).
\end{equation}
where $\gamma$ is the Lorentz factor of the particle, $v$ is the particle velocity. 
We write velocity into two parts: the equilibrium part $v_0$ and the perturbation part $\delta v$. Only the latter is considered in the Fourier space, where all the parameters are denoted in ``tilde''.

(i) We first consider that $E_\parallel$ is a constant. 
We perform Fourier transformation to Eq. (\ref{motion})
\begin{equation}
m_e\gamma^3i\left(\omega-s\frac{d\theta}{dt}\right)\delta\Tilde{v}=e\Tilde E_\theta+E_\parallel\delta(\omega)\delta(s)\delta(k_z),
\end{equation}
where the Fourier transformation to the constant $E_\parallel$ is denoted in terms of delta functions, which suggests that this term is zero when $\omega,s$ and $k_z\neq0$. One can see the presence of $E_\parallel$ cannot influence the result of $E_\theta$ obtained by \cite{Gil04}, since only the perturbation terms are considered and since all the derivations are in the Fourier space.

(ii) Next, we consider that $E_\parallel$ is a slow-changing field, i.e. $E_\parallel\propto {\rm exp}(ik_{\parallel}z-i\omega_{\parallel}t)$. The magnetic field line is assumed to have a circular configuration with an infinite $B$ strength similar to \cite{Gil04}, we use the relation $d\theta/dt=v/r$ and obtain
\begin{equation}
\delta \Tilde v=\frac{ie(\tilde E_p+\tilde E_\parallel)}{\gamma^3 m_e\left(\omega-s\frac{v}{r}\right)}.
\end{equation}
Performing Fourier transformation to the current density continuity equation, one gets
\begin{equation}\label{j_theta}
-i\omega\delta\rho+i\frac{s}{r}\tilde j_{\rm plasma}=0.
\end{equation}
The plasma current density along $z$-axis can be solved as
\begin{equation}
\Tilde{j}_{\rm plasma}=\frac{i\omega_p^2\omega(\Tilde{E}_{\parallel,s}+\Tilde E_\theta)}{4\pi\gamma^3(\omega-s\frac{v}{r})^2}.
\end{equation}
The plasma current density is modified in the presence of $E_\parallel$ compared with \cite{Gil04}. According to the electric potential, the global electric field can be written as
\begin{equation}
\tilde E_\theta=i\left(\frac{\omega}{c}\tilde A_\theta-\frac{s}{r}\tilde \phi\right).
\end{equation}
We apply the short wave approximation, i.e. we simplify the current density and electric field equations in a narrow region $r-r_0\ll r_0$ in the cylindrical coordinate system and use the short wave approximation $s\gg1$ and $\omega\gg k_z c$ (see also \cite{Gil04}). The global electric field can be expressed as
\begin{equation}
\tilde E_\theta=\frac{ics^{4/3}}{\omega r_0^2}\left[2(x+a)\tilde A_{\theta}+is^{1/3}\tilde A_r'\right],
\end{equation}
where $a=s^{2/3}k_z^2c^2/2\omega^2$ is a dimensionless parameter. In order to calculate $\tilde E_\theta$, we need to find the relations of the vector potential components. The $\theta$-direction (along background magnetic field) component of wave equation in terms of vector potential can be calculated under short wave approximation as
\begin{equation}
A_\theta''+s^{-2/3}A_\theta'+2xA_\theta-2is^{1/3}A_r=-\frac{4\pi}{c}r_0^2s^{-4/3}\tilde j_{\rm plasma},
\end{equation}
where $x=(r-r_0)s^{2/3}/{r_0}$. The point-like charge current density is 
\begin{equation}
\tilde j_{\rm bunch}=\frac{QV}{R}\delta(r-R)\delta\left(\theta-\frac{Vt}{R}\right)\delta(z),
\end{equation}
where $Q$ is the net charge of the bunch and $V$ is the velocity.
The plasma current density term can be expressed in the wave equation as
\begin{equation}
\begin{aligned}
-4\pi r_0^2 s^{-4/3}\Tilde{j}_{\rm plasma}&=-i\frac{\omega_p^2\gamma\omega r_0^2}{\pi s^2c^2}(1+2s^{-2/3}\gamma^2x)^{-2}\\
&\times\left[\Tilde{E}_{\parallel,s}+i\frac{c}{\omega}\left(2\frac{s^{4/3}}{r_0^2}x+k_z^2\right)A_\theta-\frac{c}{\omega}\frac{s^{5/3}}{r_0^2}A_r'\right]\\
&=8\frac{\omega_p^2\gamma}{\omega^2}[1+2s^{-2/3}\gamma^2(a+x)]^{-2}\\
&\times\left[-i\frac{\omega}{c}\Tilde{E}_{\parallel,s}+2(x+a)A_\theta+is^{1/3}A_r'\right],
\end{aligned}
\end{equation}
where $A_r$ and $A_\theta$ are the vector potential components along radial and azimuthal directions, respectively.
We consider that the ratio of plasma frequency to FRB wave frequency is very large, i.e. $\omega_p^2\gamma/\omega^2\gg1$, then the last term in the square brackets should be small, i.e. approximately equal to zero. Then we can obtain the relation\footnote{The first factor on the right hand side can be simplified as $(1-\gamma^2/\Gamma^2)^{-2}$, which means when $\gamma\gg\Gamma$, the relation is not valid. Therefore, all the calculations are only valid when the plasma Lorentz factor is smaller than the bunch Lorentz factor.}
\begin{equation}
A_\theta=\frac{-is^{1/3}A_r'+i\omega\Tilde{E}_{\parallel,s}/c}{2(x+a)}.
\end{equation}
In order to find the matching condition of $A$ to calculate $\tilde E_\theta$, we have
\begin{equation}
\begin{aligned}
A_\theta'&=-\frac{is^{1/3}}{2(x+a)}A_r''+\frac{is^{1/3}}{2(x+a)^2}A_r'-\frac{i\omega_{\rm aw}\Tilde{E}_{\parallel,s}}{2c(x+a)^2}+\frac{i\omega_{\rm aw} k_{\rm aw}\Tilde{E}_{\parallel,s}}{2c(x+a)}\\
&\simeq-\frac{is^{1/3}}{2(x+a)}A_r''+\frac{is^{1/3}}{2(x+a)^2}A_r'\simeq-i\frac{s^{1/3}}{2(x+a)}A_r''-\frac{1}{(x+a)}A_\theta,
\end{aligned}
\end{equation}
where we ignore the last two terms on the right hand side since $\omega_{\rm aw}$ and $k_{\rm aw}$ have small magnitudes. Therefore, the slow-changing parallel electric field would not modify the final global electric field of the system.
From the $r$-direction component of the wave equation 
\begin{equation}
A_r''=-2xA_r+2is^{-1/3}A_\theta,
\end{equation}
we have
\begin{equation}
\begin{aligned}
A_\theta'&=is^{1/3}\frac{x}{(x+a)}A_r+\frac{1}{(x+a)}A_\theta-\frac{1}{(x+a)}A_\theta\\
&=is^{1/3}\frac{x}{(x+a)}A_r.
\end{aligned}
\end{equation}
The second derivative of $A_2$ is calculated as
\begin{equation}
A_\theta''=is^{1/3}\left[\frac{1}{(x+a)}A_r-\frac{x}{(x+a)^2}A_r+\frac{x}{(x+a)}A_r'\right].
\end{equation}
Thus the $\theta$-direction component of wave equation can be re-written and simplified as
\begin{equation}
\begin{aligned}
A_\theta''+2xA_\theta+2is^{-1/3}
&=\frac{is^{1/3}A_r}{(x+a)}-\frac{is^{1/3}xA_r}{(x+a)^2}+2is^{-1/3}A_r\\
&=is^{1/3}\left[\frac{a}{(x+a)^2}+2is^{-2/3}\right]A_r.
\end{aligned}
\end{equation}
With the point-like charge current density, we can obtain a general expression of the wave equation along the $\theta$-direction
\begin{equation}
\begin{aligned}
is^{1/3}\left[\frac{a}{(x+a)^2}+2is^{-2/3}\right]&A_r=-\frac{\omega_p^2\gamma}{\omega^3}\frac{i8s^{2/3}\tilde E_\theta}{[1+2s^{-2/3}\gamma^2(a+x)]^2}\\
&-8\pi^2QVs^{-2/3}\delta(x-X)\delta(\omega-s\frac{V}{R}),
\end{aligned}
\end{equation}
where $X=-a-s^{2/3}/2\Gamma^2$ denotes the position of the emitting particle.
One can notice that the second term on the right hand side is smaller than the first term since $s\gg1$, thus $\tilde E_\theta$ can be calculated as
\begin{equation}\label{Ep}
\tilde E_\theta=-\frac{1}{8}s^{-1/3}\left(\frac{\omega_p^2\gamma}{\omega^3}\right)^{-1}[1+2s^{-2/3}\gamma^2(a+x)]^2\frac{aA_r}{(x+a)^2}.
\end{equation}
Under the short wave approximation we can obtain two relations:
\begin{equation}
[1+2s^{-2/3}\gamma^2(a+X)]^2=\left(1-\frac{\gamma^2}{\Gamma^2}\right)^2,
\end{equation}
and
\begin{equation}
\frac{a}{(X+a)^2}=2s^{-2/3}\Gamma^4\frac{k_z^2}{\omega^2}.
\end{equation}
We apply the above two relations and use $\omega R\approx sc$ to simplify Eq.(\ref{Ep}) as
\begin{equation}
\tilde E_\theta=2i\pi^2s^{-1}\Gamma^4\frac{k_z^2c}{\omega_p^2}\frac{Q}{\gamma}\frac{(1-\frac{\gamma^2}{\Gamma^2})^2}{1+\frac{k_z^2\Gamma^2}{\omega^2}}\delta\left(s-\omega\frac{R}{V}\right)u,
\end{equation}
where $u$ is a  dimensionless function of the $\theta$-direction component of the wave equation. This result is consistent with \cite{Gil04} (see their Eq.(32)). We can then conclude that the slow-changing parallel electric field would not modify the expression of the suppression factor for the $E_\theta$ part. Therefore, we can linearly add the $E_\parallel$ in the calculation of the emission power.




\bsp	
\label{lastpage}
\end{document}